\documentclass[]{an}
\usepackage{graphicx}
\usepackage{times}

\def\gsim{\lower.4ex\hbox{$\;\buildrel >\over{\scriptstyle\sim}\;$}} %$
\def\lsim{\lower.4ex\hbox{$\;\buildrel <\over{\scriptstyle\sim}\;$}} %$

\def\[{\begin{eqnarray}}
\def\]{\end{eqnarray}}

\begin{document}

% The following seven commands are intended for editorial usage and should be ignored by
% the author(s).
 \Pagespan{1}{}% Document's page range.
% If second parameter is left empty, the last page is computed automatically.
 \Yearpublication{2006}
 \Yearsubmission{2005}
 \Month{11}%
 \Volume{999}%
 \Issue{88}%
% \DOI{This.is/not.aDOI}%

\title{Diamagnetic pumping near the base of a stellar convection zone}

\author{L.\,L.~Kitchatinov\inst{1,2}\fnmsep\thanks{Corresponding author:
 kit@iszf.irk.ru}
  \and G.~R\"udiger\inst{1}
}
 \titlerunning{Diamagnetic pumping near the base of a stellar convection zones}
 \authorrunning{L.\,L.~Kitchatinov \& G.~R\"udiger}
 \institute{ Astrophysikalisches Institut Potsdam, An der Sternwarte
             16, D-14482 Potsdam, Germany
 \and Institute for Solar-Terrestrial Physics, P.O.~Box 291,
      Irkutsk 664033, Russia}

 \received{}
 \accepted{}
 \publonline{later}

 \keywords{instabilities --
                      stars: interiors --
           stars: magnetic fields --
           Sun: magnetic field}

 \abstract{%
The property of inhomogeneous turbulence in conducting fluids to
expel large-scale magnetic fields in the direction of decreasing
turbulence intensity is shown as important for the magnetic field
dynamics near the base of a stellar convection zone. The downward
diamagnetic pumping confines a fossil  internal magnetic field in
the radiative core so that the field geometry is appropriate for
formation of the solar tachocline. For the stars of solar age, the
diamagnetic confinement is efficient only if the ratio of turbulent
magnetic diffusivity $\eta_{\rm T}$ of the convection zone to the
(microscopic or turbulent) diffusivity $\eta_{\rm in}$ of the
radiative interiour is $\eta_{\rm T}/\eta_{\rm in} \geq 10^5$.
Confinement in younger stars require larger $\eta_{\rm T}/\eta_{\rm
in}$. The observation  of persistent magnetic structures on young
solar-type stars can thus provide evidences for the nonexistence  of
tachoclines in stellar interiors and on the level of turbulence in
radiative cores.  }

\maketitle

%%%%%%%%%%%%%%%%%%%%%%%%%%%%%%%%%%%%%%%%%%%%%%%%%%%%%%%%%%%%%%%%%%%%%%%
\section{Introduction}\label{introduction}
%%%%%%%%%%%%%%%%%%%%%%%%%%%%%%%%%%%%%%%%%%%%%%%%%%%%%%%%%%%%%%%%%%%%%%%
This paper concerns diamagnetism of stellar convective turbulence to
show that the diamagnetic pumping can be important for formation of
the solar tachocline.

Turbulent conducting fluids are known to expel large-scale magnetic
fields in the direction of decreasing turbulent intensity. This
effect of turbulent diamagnetism has been first found for
inhomogeneous 2D turbulence (Zeldovich \cite{Z57}). For this case,
the effective velocity of magnetic field transport is proportional
to the gradient of the turbulent magnetic diffusivity, ${\vec
U}\!_\mathrm{dia} = -{\vec\nabla} \eta_\mathrm{T}$. The minus sign
in the right shows the sense of turbulent magnetism: it is not para-
but dia-magnetism so that magnetic fields are pushed away from the
regions of relative high turbulent intensity. The diamagnetism is
closely related to the magnetic field expulsion from regions of
circular motion (Weiss \cite{W66}). It was detected in 3D numerical
simulations (Brandenburg et al. \cite{Bea96}; Tobias et al.
\cite{Tea98}, \cite{Tea01}; Dorch \& Nordlund \cite{DN01}; Thomas et
al. \cite{Tea02}; Ziegler \& R\"udiger \cite{ZR03}) and laboratory
experiments with turbulent liquid sodium (Spence et al.
\cite{Sea07}). For 3D nearly isotropic turbulence the expression for
the diamagnetic transport velocity becomes
\begin{equation}
   {\vec U}\!_\mathrm{dia} =-\frac{1}{2}{\vec\nabla}\eta_\mathrm{T}
   \label{1}
\end{equation}
(Krause \& R\"adler \cite{KR80}). The influence of rotation and
magnetic fields produce anisotropy and quenching of both turbulent
pumping and diffusion (Kitchatinov \cite{K88}; Kitchatinov \&
R\"udiger \cite{KR92}). In particular the strong magnetic quenching
of the diamagnetism $\sim B^{-3}$ for super-equipartition fields has
been found. The isotropic parts of the field advection and diffusion
are still related by the Eq.~(\ref{1}) when rotation or strong
magnetic field is imposed. Extensive numerical simulations -- also
as a test for the early theoretical SOCA-expressions -- are due to
Ossendrijver et al. (\cite{OSBR02}) and K\"apyl\"a et al.
(\cite{Kea06}).
\begin{figure}
    \includegraphics[width=8.cm, height=4.7cm]{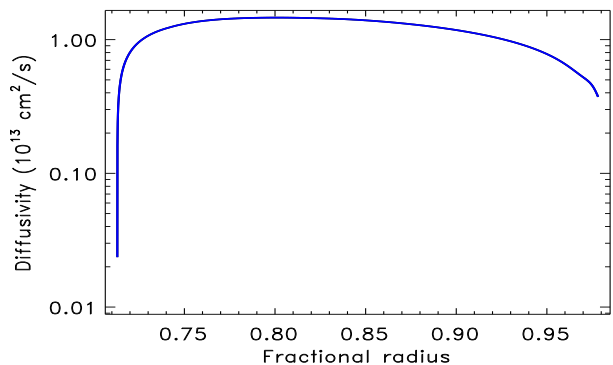}
    \caption{The radilal profile of the turbulent magnetic diffusivity
             in the solar convection zone
             after the   model of Stix \& Skaley
              (\cite{SS90}). The
             diamagnetic pumping should be very strong near the base of the convection
             zone where the diffusivity  almost jumps by orders of magnitude.
              }
    \label{f1}
\end{figure}

The turbulent pumping can be significant in various astrophysical
contexts. Its strong effect is known in models of the galactic
dynamo where the pumping may exceed the azimuthal $\alpha-$effect so
that the dynamo can even switch off. In the present paper we suggest
its importance also for the Sun and solar-type stars with external
convection zones where it can participate in formation of the
tachocline immediately below the base of the convection zone.

Figure~\ref{f1} shows the radial profile of the magnetic diffusivity
in the solar convection zone estimated with the mixing-length
relation $\eta_\mathrm{T} = u'\ell/3$. The diffusivity varies
sharply near the base of convection zone. Its gradient represents a
velocity of the  downward diamagnetic pumping (\ref{1}). The
amplitude of the velocity is up to 50 m/s  exceeding both the
amplitude of the $\alpha$-effect (1--10 m/s)  and the meridional
flow in the convection zone ($\lsim 10$ m/s).

The tachocline can be explained by an effect of a weak internal
magnetic field of the solar radiative core (R\"udiger \& Kitchatinov
\cite{RK97}, \cite{RK07}). The magnetic tachocline theory does not
strongly restrict  the magnetic field amplitude. Even a weak field
well below 1~Gauss can produce the tachocline. The tachocline theory
is, however, very sensitive to the field structure. The internal
field should be almost totally confined inside the radiative core in
order to produce the tachocline (MacGregor \& Charbonneau
\cite{MC99}; Kitchatinov \& R\"udiger \cite{KR06}). The downward
diamagnetic pumping near the  convection zone base can keep the
internal magnetic field confined in the radiative core thus
providing the field geometry necessary for the tachocline formation
(Garaud \cite{G07}; Garaud \& Rogers \cite{GR07}). The diamagnetic
confinement is estimated in this paper almost linear and in large
scales. The main astrophysical question is how long the turbulent
pumping would need to produce a confined magnetic field from an
opened field configuration. Note that the diffusion time in the
solar interior is longer by many order of magnitudes than the
diffusion time in the convection zone. If the time to produce a
confined magnetic geometry would, e.g., equal 10 Myr then  -- if the
tachocline is magnetic by origin -- the tachocline would only exist
in older stars. On the other hand, the confinement mechanism via
meridional flow with amplitudes of order 10 m/s works much faster
(Kitchatinov \& R\"udiger \cite{KR06}). There is thus the hope that
future asteroseismology observations will decide which of both the
mechanisms is really acting.

Also the evolution of  external fields in the convection zone under
the presence of the diamagnetic effect must be reconsidered. This
process may have consequences for the overshoot dynamo models (see
Gilman \cite{G92}; Belvedere, Lanzafame \& Proctor \cite{BLP91};
R\"udiger \& Brandenburg \cite{RB95}) although much stronger
dynamo-fields may suppress turbulent advection of the field.

%%%%%%%%%%%%%%%%%%%%%%%%%%%%%%%%%%%%%%%%%%%%%%%%%%%%%%%%%%%%%%%%%
\section{Equations}
%%%%%%%%%%%%%%%%%%%%%%%%%%%%%%%%%%%%%%%%%%%%%%%%%%%%%%%%%%%%%%%%%
Consider the magnetic field in a sphere with an inner core where
magnetic diffusivity $\eta_{\rm in}$ is relative low and an outer
spherical shell with the diffusivity $\eta_{\rm T} \gg \eta_{\rm
in}$. The diffusivity varies smoothly  between these two values
around the interface at $r = R_\mathrm{in}$,
\begin{equation}
   \eta = \eta_{\rm in} + \frac{1}{2}\left(\eta_{\rm T} - \eta_{\rm in}\right)
          \left( 1 + \mathrm{erf}\left(\frac{r-R_\mathrm{in}}
          {h_\mathrm{d}}\right)\right),
   \label{2}
\end{equation}
where erf is the error function and $h_\mathrm{d}$ defines the width
of the transition layer. The evolution of the large-scale magnetic
field is governed by the diffusion equation
\begin{equation}
   \frac{\partial{\vec B}}{\partial t} =-{\vec\nabla}\times
   \left(\sqrt{\eta}{\vec\nabla}\times\left(\sqrt{\eta}{\vec B}
   \right)\right) ,
   \label{3}
\end{equation}
which includes the diamagnetic transport with the effective velocity
(\ref{1}). A vacuum boundary condition is applied at the top, $r =
R$.

Equation~(\ref{3}) describes the decay of the field. We expect the
decay being sufficiently slow for the internal fields to survive in
the stellar radiation cores on  time scales of Gyrs (Cowling
\cite{C45}). With the diamagnetism included the  magnetic fields are
expected to be confined in the core. A solution of the eigenvalue
problem
\begin{equation}
    {\vec B}({\vec r},t) = \exp (\sigma t){\vec b}({\vec r}) ,
    \label{4}
\end{equation}
provides both the decay time $\tau = -\sigma^{-1}$ and the field
geometry. The eigenmodes with the smallest decay rate is the most
significant one. The degree of confinement of an axisymmetric
poloidal field,
\begin{equation}
    {\vec b} = {\vec\nabla}\times\left({\vec e}_\phi\frac{A}{r\sin\theta}\right) ,
    \label{5}
\end{equation}
can be described by the escape parameter
\begin{equation}
    \delta\phi = \frac{\mathrm{max}|A|_{r = R_\mathrm{in}}}
    {\mathrm{max}|A|_{r \leq R_\mathrm{in}}}
    \label{6}
\end{equation}
(Kitchatinov \& R\"udiger \cite{KR06}). Here,  $r,\theta ,\phi$ are
standard spherical coordinates, and ${\vec e}_\phi$ is the azimuthal
unit vector. The parameter (\ref{6}) measures the ratio of the
characteristic values of magnetic flux through the interface to the
flux in the core. The maximum value of $\delta\phi = 1$ corresponds
to an open field structure. The smaller $\delta\phi$ the more
confined to the core the internal field is. A tachocline can be
formed by the internal field if the escape parameter is sufficiently
small, i.e.
\begin{equation}
   \delta\phi \lsim 10^{-2},
   \label{delta}
\end{equation}
(R\"udiger \& Kitchatinov \cite{RK07}).

The eigenvalue problem for the potential $A$ of the poloidal field
formulated in terms of the spherical harmonics, $A = A_l(r)
P^1_l(\cos\theta )\sin\theta$, leads to the equation
\begin{equation}
    \sigma A_l = \sqrt{\eta}\frac{\mathrm{d}}{\mathrm{d} r}\left(
    \sqrt{\eta}\frac{\mathrm{d} A_l}{\mathrm{d} r}\right)
    - \eta \frac{l(l+1)}{r^2}\ A_l
    \label{7}
\end{equation}
with the vacuum boundary condition
\begin{equation}
    \frac{\mathrm{d} A_l}{\mathrm{d} r} = -\frac{l}{R}\ A_l
    \label{8}
\end{equation}
at $r=R$.

We can  solve  Eq.~(\ref{7})  numerically for the  continuous
diffusivity profile (\ref{2}).  But also an analytical solution can
be found for a discontinuous profile of $\eta$
($h_\mathrm{d}\rightarrow 0$). In this case the solution of
Eq.~(\ref{7}) for constant $\eta$ can be used, i.e.
\begin{equation}
   A_l = a \sqrt{r}\ J_{l+\frac{1}{2}}\left(\frac{r}{\sqrt{\eta\tau}}\right)
          + b \sqrt{r}\ J_{-l-\frac{1}{2}}\left(\frac{r}{\sqrt{\eta\tau}}\right),
   \label{9}
\end{equation}
where $J_\nu$ is the Bessel function, $\tau = -\sigma^{-1}$ is the
inverse eigenvalue, and $a$ and $b$ are  free constants. Solutions
with different sets of constants $a$ and $b$ apply to the inner core
and the envelope. The boundary conditions
\begin{equation}
   \left[ A\right] = 0, \ \ \
   \left[\sqrt{\eta}\frac{\mathrm{d} A}{\mathrm{d} r}\right] = 0\ \ \ \
   \mathrm{at}\ \ r = R_\mathrm{in}
   \label{10}
\end{equation}
can be applied at the interface. In the inner core, $b$  must be
zero to avoid a singularity at $r = 0$. The other constant $a$ can
freely be chosen, e.g. $a = 1$, to normalize the linear solution.
Then, Eqs. (\ref{8}) and (\ref{10}) provide three conditions to
define both the constants $a$ and $b$ in the outer shell and the
eigenvalue.

%%%%%%%%%%%%%%%%%%%%%%%%%%%%%%%%%%%%%%%%%%%%%%%%%%%%%%%%%%%%%%%%%
\section{Results and discussion}
%%%%%%%%%%%%%%%%%%%%%%%%%%%%%%%%%%%%%%%%%%%%%%%%%%%%%%%%%%%%%%%%%
\subsection{Normal modes of the internal field}
\begin{figure}[htb]
    \includegraphics[width=8.0cm,height=6.0cm]{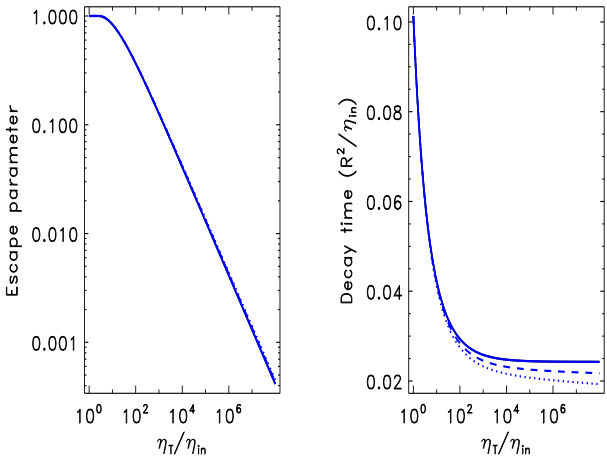}
    \caption{Dependence of the escape parameter (\ref{6}) and the decay
    time of the most long-living dipolar eigenmode on the diffusivity
    contrast $\eta_{\rm T}/\eta_{\rm in}$. Full line corresponds to analytical
    solution for discontinuous change of $\eta$. Dashed and dotted lines show
    the numerical results for smooth profiles (\ref{2}) with
    $h_\mathrm{d} = 0.01R$ and $h_\mathrm{d} = 0.02R$ respectively.
    All three lines overlap in the left panel.
              }
    \label{f2}
\end{figure}

\begin{figure}[htb]
    \includegraphics[width=2.7cm]{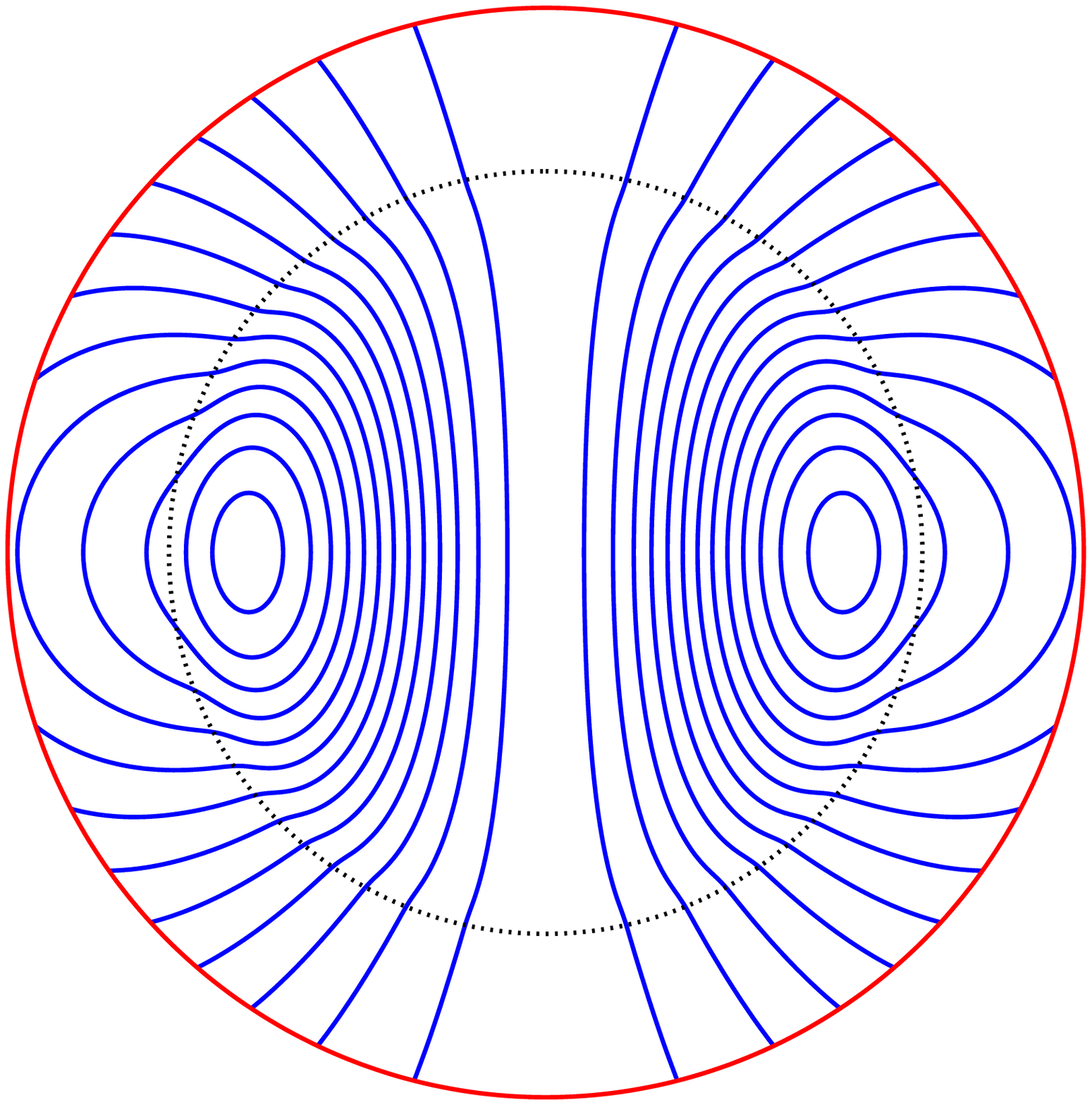}
    \includegraphics[width=2.7cm]{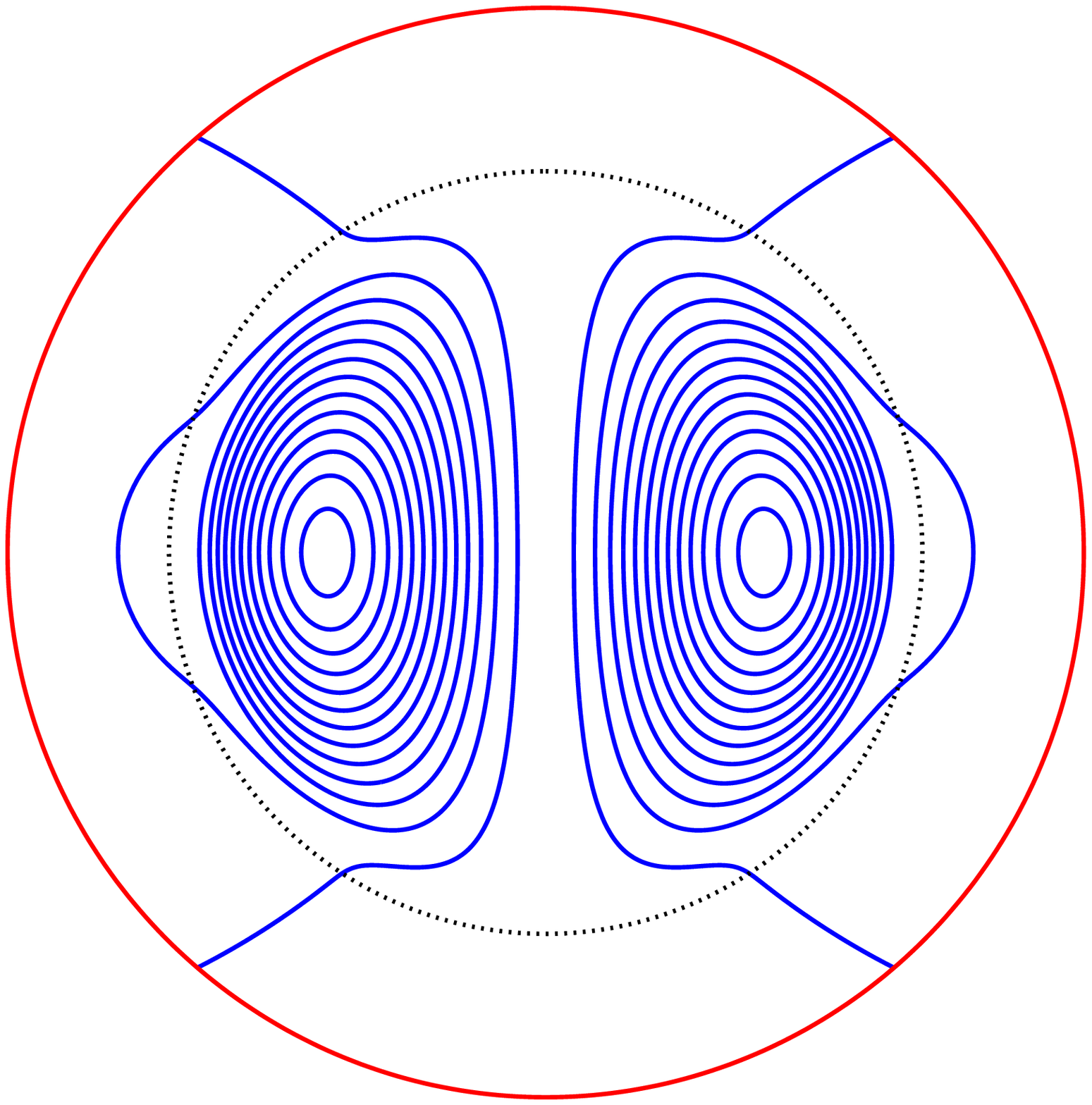}
    \includegraphics[width=2.7cm]{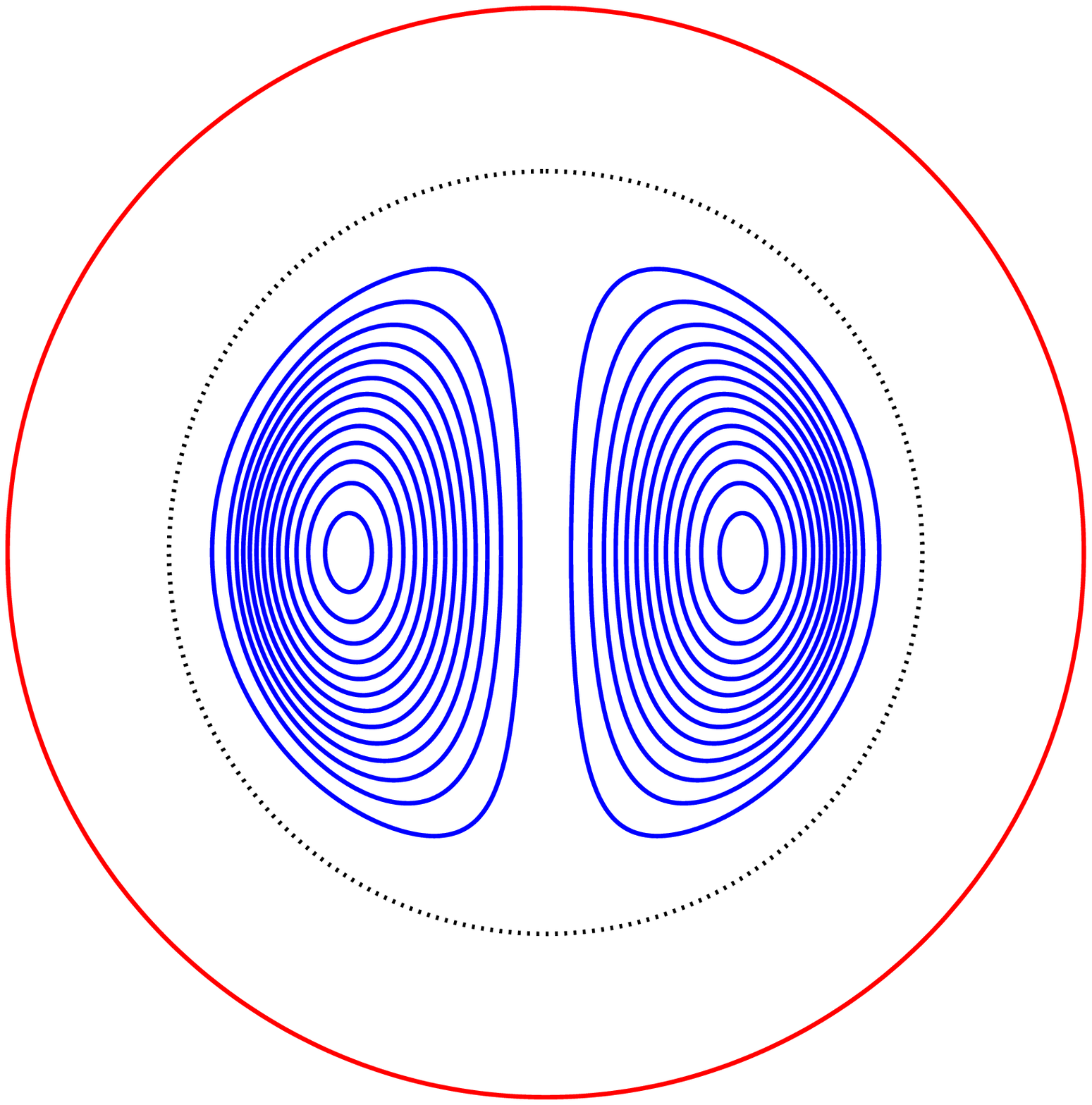}
    \caption{Poloidal field eigenmodes. The diffusivity contrast varies as
    $\eta_{\rm T}/\eta_{\rm in} = 10, 10^3, 10^5$ from the left to the right. The dotted
        circle shows the interface $r = R_\mathrm{in}$.
              }
    \label{f3}
\end{figure}

%%%%%%%%%%%%%%%%%%%%%%%%%%%%%%%%%%%%%%%%%%%%%%%%%%%%%%%%%%%%%%%%%
Figure~\ref{f2} shows the dependencies of the escape parameter
(\ref{6}) and the decay time of the slowest decaying eigenmode of
dipolar symmetry as a function of  the diffusivity contrast
$\eta_{\rm T}/\eta_{\rm in}$ between the inner core and outer
envelope. The decay times are given separately for the analytical
solution and numerical solutions for two values of transition width
$h$. The trapping  of the field  in the radiative core grows with
the diffusivity contrast. Figure~\ref{f3} shows the structure of the
eigenmodes.

The tachocline can be formed by the internal field if the escape
parameter is sufficiently small (\ref{delta}). After Fig.~\ref{f2},
this is the case if $\eta_{\rm T}/\eta_{\rm in} > 10^5$. For an eddy
diffusivity $\eta_{\rm T} \sim 10^{13}$ cm$^2$s$^{-1}$ in the
convection zone, this inequality means that
\begin{equation}
   \eta_{\rm in} < 10^8\ \mathrm{cm}^2\mathrm{s}^{-1}
   \label{11}
\end{equation}
in the tachocline. Hence, the tachocline can be only mildly
turbulent if it is magnetic by origin.

The internal field can also be confined by a global meridional flow
penetrating the radiative core from convection zone (Kitcha\-tinov
\& R\"udiger \cite{KR06}). This type of confinement also can be
efficient only if $\eta_{\rm in}$ is sufficiently low with the same
upper bound (\ref{11}) as for the diamagnetic confinement. The
reason for this coincidence might be that the confinements by
meridional flow or inhomogeneous turbulence represent basically the
same mechanism on different spatial scales. The condition (\ref{11})
for the field expulsion from the region of motion to be efficient in
the Sun does not, however, depend on the scale at all.

%%%%%%%%%%%%%%%%%%%%%%%%%%%%%%%%%%%%%%%%%%%%%%%%%%%%%%%%%%%%%%%%%
\subsection{Internal field evolution towards a core-confined geometry}
\begin{figure}[htb]
    \includegraphics[width=8.0cm, height=8.0cm]{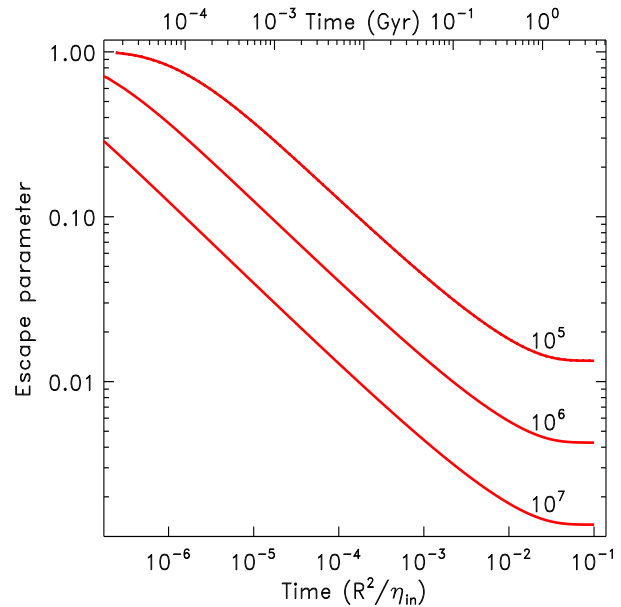}
    \caption{Escape parameter (\ref{6}) as function of time for the runs starting
    from an open field structure. The lines are marked by the corresponding
    values of the diffusivity contrast. The upper scale gives the physical time
    for the Sun computed for the microscopic diffusivity
    $\eta_{\rm in} = 3\times 10^3$~cm$^2$/s.
              }
    \label{f4}
\end{figure}

%%%%%%%%%%%%%%%%%%%%%%%%%%%%%%%%%%%%%%%%%%%%%%%%%%%%%%%%%%%%%%%%%
A freely decaying field eventually approaches the most long-living
eigenmode. To find the characteristic time scale for the field
convergence to confined geometry of the eigenmodes the field was
evolved in time with Eq.~(\ref{3}) starting from an initial field of
open structure. Figure~\ref{f4} shows the dependencies of the escape
parameter (\ref{6}) on time for several diffusivity contrasts. The
$\delta\phi$ decreases in time but eventually approaches a constant
escape parameter of the corresponding eigenmode. The confinement
proceeds on the long time scale $R^2/\eta_{\rm in}$ of the internal
diffusion despite the diamagnetic pumping is fast. Why the
confinement is so slow can be understood by observing the evolution
of the magnetic field structure.

\begin{figure*}[htb]
   \hbox
   { \includegraphics[width=4.7cm]{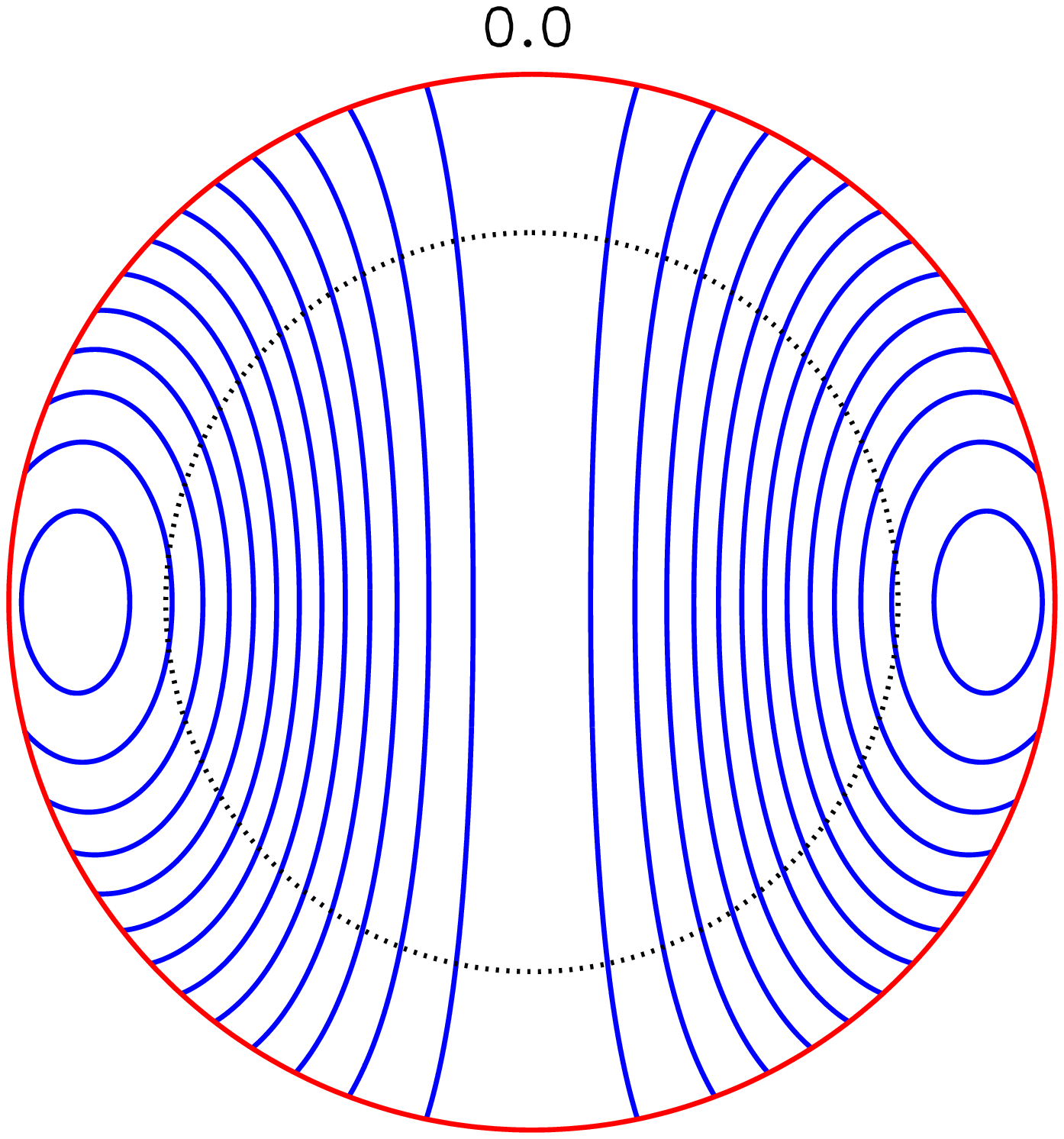}\ \ \ \ \ \ \ \ \ \ \ \ \ \ \ \
    \includegraphics[width=4.7cm]{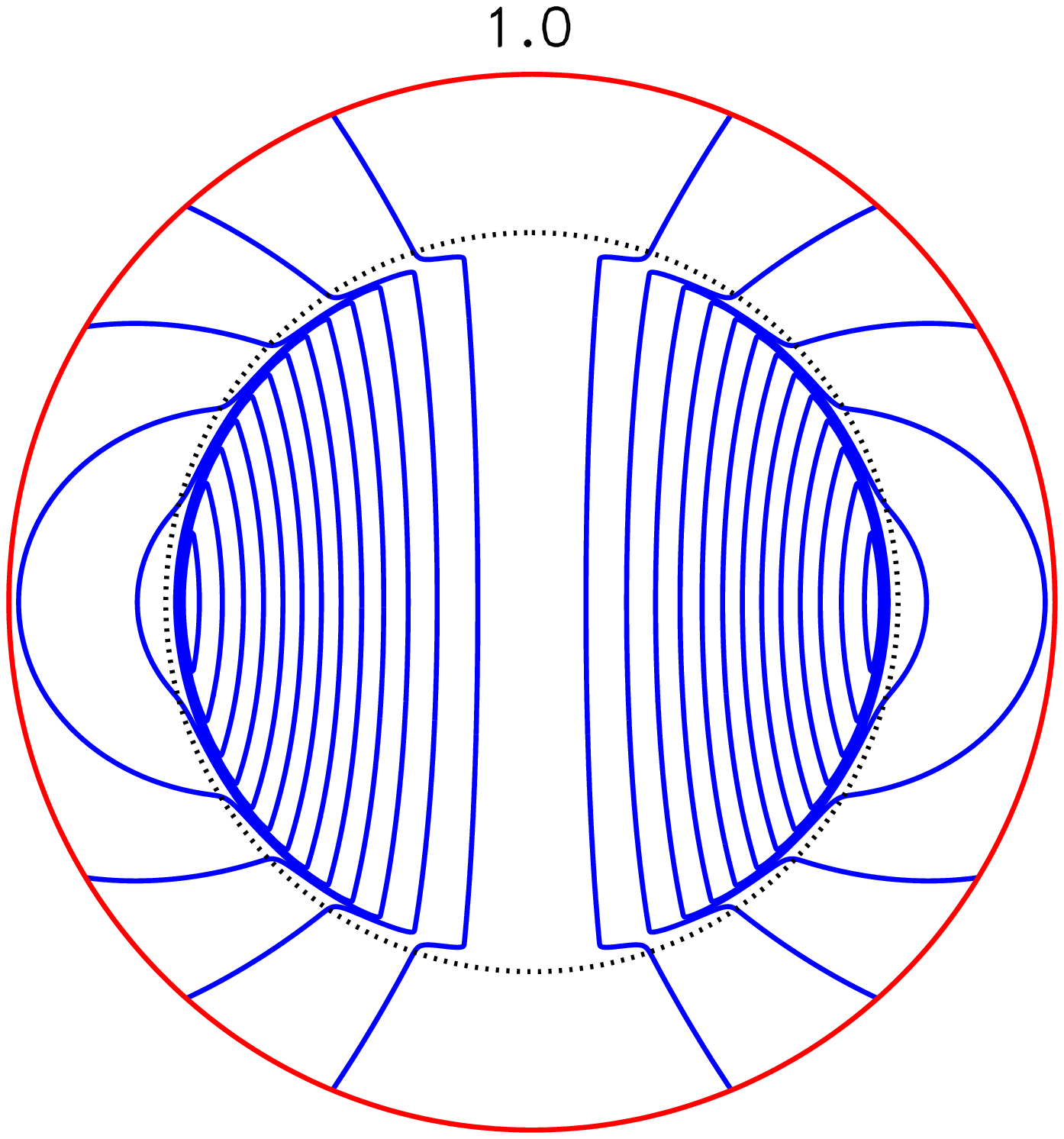}\ \ \ \ \ \ \ \ \ \ \ \ \ \ \ \
    \includegraphics[width=4.7cm]{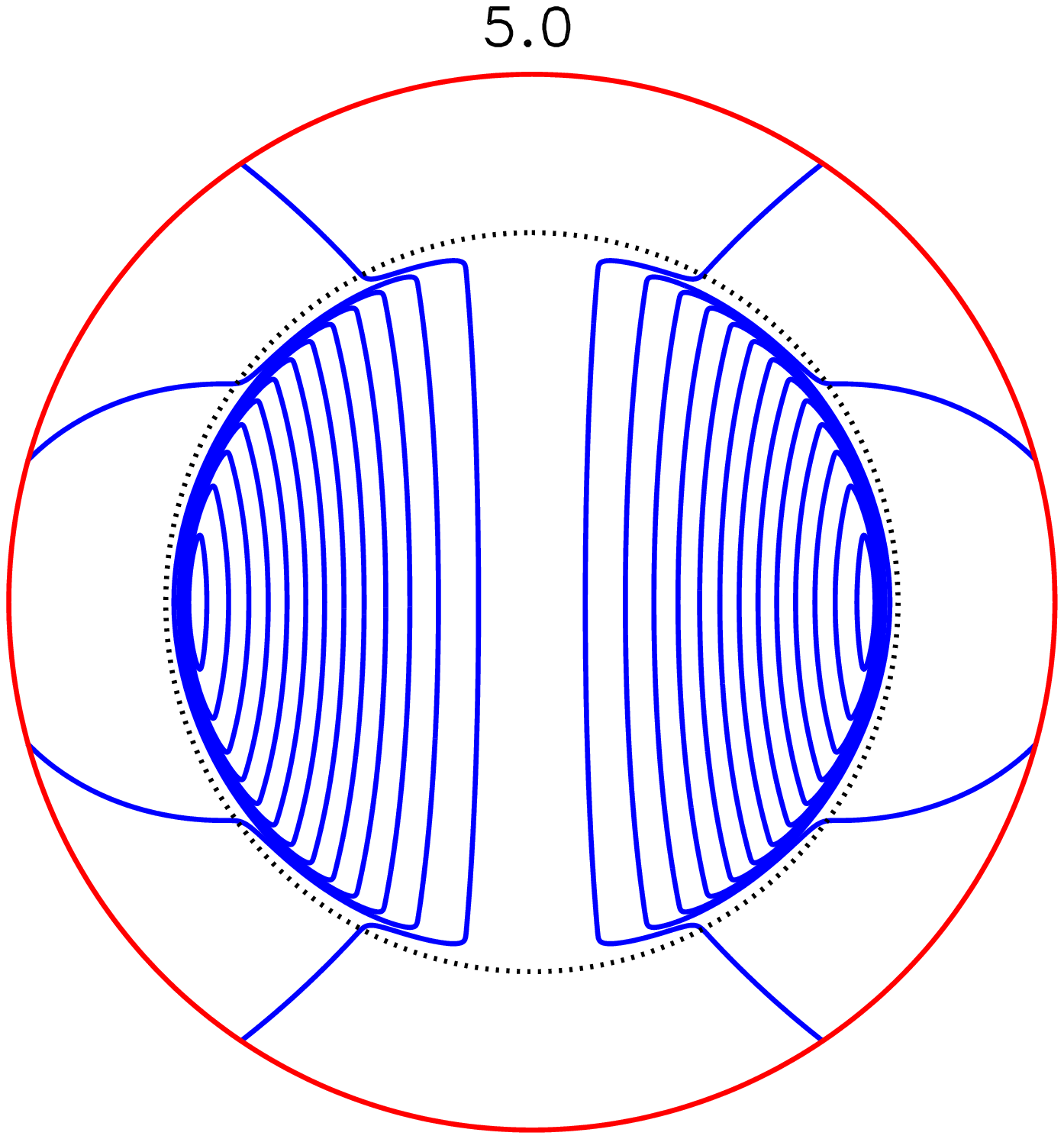}}
    \caption{Initial field (left) and the structure of the field after
    it was evolved with Eq.~(\ref{3}) to later instants (middle and right).
    Time in {\em external} diffusion units $R^2/\eta_{\rm T}$ is shown on the top,
    all for $\eta_{\rm T}/\eta_{\rm in} = 10^5$.
        }
    \label{f5}
\end{figure*}

Figure~\ref{f5} shows the initial phases  of the confinement. After
one (short) diffusion time, $R^2/\eta_{\rm T}$, the strong
concentration of the field in the upper radiative core is already
formed. The strong downward increase of the poloidal field across
the interface is already in balance with diamagnetic pumping. The
confinement slows down  after this balance is achieved. The high
concentration of the field  immediately beneath the interface does
not allow a high degree of confinement even with a strong decrease
of the field from the upper core to the lower convection zone. The
field must be further smoothed over the entire core to attain the
eigenmode structure with high confinement. This smoothing is, of
course, slow. Nevertheless, the internal fields can be confined
already in young stars of the age of several tens of million years
if the diffusivity contrast between the convection zone and the
radiative core is sufficiently high (Fig.~\ref{f4}). Our
calculations, therefore, predict that  even rather young stars may
already possess tachoclines.

Observations can probe whether the solar-type stars of young
clusters possess persistent magnetic structures which do not
considerably change for years. If the fields rooted in the radiative
core are not confined there, they will penetrate to the surface with
no remarkable variations. If persistent magnetic structures are
found then the internal field is not confined and a tachocline
cannot be formed. Simultaneously, that would indicate a certain
level of turbulence in the stellar cores.

%%%%%%%%%%%%%%%%%%%%%%%%%%%%%%%%%%%%%%%%%%%%%%%%%%%%%%%%%%%%%%%%%
%%%%%%%%%%%%%%%%%%%%%%%%%%%%%%%%%%%%%%%%%%%%%%%%%%%%%%%%%%%%%%%%%
\acknowledgements This work was supported by the Deutsche
Fo\-r\-schungs\-ge\-mein\-schaft and by the Russian Foundation for
Basic Research (project 05-02-04015).
%%%%%%%%%%%%%%%%%%%%%%%%%%%%%%%%%%%%%%%%%%%%%%%%%%%%%%%%%%%%%%%%%

%%%%%%%%%%%%%%%%%%%%%%%%%%%%%%%%%%%%%%%%%%%%%%%%%%%%%%%%%%%%%%%%%%%
\end{document}